\documentclass[referee]{raa}            % referee version: for submission
\usepackage{graphicx,times}     %for PS/EPS graphics inclusion, new
\usepackage{longtable,lscape}
\usepackage{natbib}
\usepackage{amssymb,amsmath}
\usepackage{multirow}
\usepackage{graphicx}
\usepackage{latexsym}
\usepackage{bm}
\usepackage{amssymb,amsmath} % needed for $\mathcal{R}$
\bibpunct{(}{)}{;}{a}{}{,}

\usepackage[a4paper=true,dvipdfm=true,pagebackref=true]{hyperref}
\hypersetup{colorlinks = true, linkcolor = green, anchorcolor = red, citecolor = blue, filecolor = red, pagecolor = red, urlcolor = red}

\begin{document}

  \title{$L_{\rm syn}-E_{\rm syn, p}-\delta$ relation in Active Galactic Nucleus Jets and Implication for the physical origin of the $L_{\rm p}-E_{\rm p,z}-\Gamma_0$ relation of Gamma-Ray Bursts
%\footnotetext{$*$ Supported by the National Natural Science Foundation of China.}
}
%   \subtitle{I. Place Your Subtitle Here}
 % \subtitle{}

  \volnopage{Vol.0 (20xx) No.0, 000--000}      %%preserved for Editor. DOn't remove!
  \setcounter{page}{1}          %%starting page, preserved for Editor. DOn't remove!

  \author{Xiao-Li Huang
     \inst{1,2}
  \and En-Wei Liang
     \inst{2}
   }

    \institute{School of Astronomy and Space Science, Nanjing University, Nanjing 210023, China;\\
        \and
             Guangxi Key Laboratory for Relativistic Astrophysics, School of Physical Science and Technology, Guangxi University, Nanning 530004, China; lew@gxu.edu.cn\\
\vs\no
   {\small Received~~20xx month day; accepted~~20xx~~month day}}

\abstract{ High energy photon radiations of gamma-ray bursts (GRBs) and active galactic nuclei (AGNs) are dominated by their jet radiations. We examine wether the synchrotron radiations of jets in BL Lacs, flat spectrum radio quasars (FSRQs), and Narrow Line Seyfert 1 galaxies (NLS1s) follow the relation between the prompt gamma-ray emission and the initial Lorentz factor ($\Gamma_0$) of GRBs. It is showed that the AGNs sample does not agree with the $L_{\rm p}-E_{\rm p,z}-\Gamma_{0}$ of GRBs. In addition, we obtain a tight relation of $L_{\rm syn}\propto E^{0.45\pm0.15}_{\rm syn,p}\delta^{3.50\pm0.25}$ for FSRQs and NLS1 galaxies, where $L_{\rm syn}$ is the luminosity at peak photon energy $E_{\rm syn, p}$ of the synchrotron radiations. This relation is different from the $L_{\rm p}-E_{\rm p,z}-\Gamma_{0}$ relation of GRBs. The dependence of $L_{\rm syn}$ to $\delta$ is consistent with the expectation of the Doppler boosting effect for the FSRQs and NLS1 galaxies, but it is not for GBRs. We argue that $\Gamma_0$ may be a representative of the kinetic power of the radiating region and the tight $L_{\rm p}-E_{\rm p, z}-\Gamma_0$ relation is shaped by the radiation physics and the jet power together.
\keywords{Gamma-Ray Bursts; Active Galactic Nuclei; Relativistic Jets; Non-thermal Radiation}}

   \authorrunning{Huang \& Liang}            %author_head in even pages
   \titlerunning{Jet Radiations of GeV-TeV AGNs and GRBs}  % title_head in odd pages

   \maketitle
%% The author head (on even pages) and the title head (on odd pages) will be
%% automatically extracted from \author{} and \title{}. Whenever the title is too long,
%% you will be asked to supply a shorter one by inserting either \authorrunning{} or
%% \titlerunning{} before \maketitle. Anyway, you can specify your own heads.
%%
%%
%% Note: In the following text body of your manuscript, please note several differences from
%%       other major journals:
%% (1) \subsection{Please Capitalize the First Letter of Each Notional Word in Subsection Title}
%% (2) Please Capitalize the First Letter of Each Notional Word in all tables' captions

\section{Introduction}

Relativistic jets are ubiquitous in the Universe and have been detected in a very diverse range of black hole (BH) systems ranging from stellar mass to supermassive scale. It is believed that gamma-ray bursts (GRBs) are produced by an ultra-relativistic jet powered by stellar black holes from core collapses of massive stars (e.g. \citealt{Woosley+1993}) or mergers of two compact stars (e.g. \citealt{Eichler+1989,Paczynski1991, Kumar+Zhang+2015}), and the high energy photon radiations of active galactic nuclei (AGNs) are dominated by radiations from a mildly relativistic jet fed by accretion of their central super-massive BHs (\citealt{Urry+Padovani+1995, Ghisellini+etal+2009, Zhang+etal+2012, Zhang+etal+2015, Liang+etal+2015, Sun+etal+2015, Zhu+etal+2016}).

The observed radiations from a jet are boosted by the Doppler effect in case of the jet toward the earth. It is generally believed that typical GRBs and blazars, including flat spectrum radio quasars (FSRQs) and BL Lacs, as well as GeV-selected Seyfert 1 galaxies, are on-axis or small angle off-axis observed to their jets (\citealt{Urry+Padovani+1995, Kumar+Zhang+2015, Sun+etal+2015}). Therefore, the Doppler boosting factor ($\delta$) is mainly dependent on the the Lorentz factor ($\Gamma$) of a relativistic jet. There are three methods to estimate the initial Lorentz factor ($\Gamma_0$) of a GRB fireball. The first one is to use the fireball deceleration time derived from the onset peaks observed in early optical afterglow lightcurves (\citealt{Sari+Piran+1999, Kobayashi+etal+1999, Liang+etal+2010}). The second one is based on the "compactness" argument by analysing the high energy spectral cutoffs or breaks of the prompt emission of GRBs (\citealt{Lithwick+Sari+2001, Tang+etal+2015}). The third method is to use the photosphere radiation in some GRBs (\citealt{Pe'er+etal+2007, Peng+etal+2014, Zou+etal+2015}). By deriving the $\Gamma_0$ values with the first method for a sample of GRBs, \cite{Liang+etal+2010} discovered a relation between $\Gamma_{0}$ and the isotropic gamma-ray energy $E_{\rm iso}$ of GRBs. \cite{Lv+etal+2012} showed that the isotropic luminosity $L_{\rm iso}$ also depends on $\Gamma_{0}$.

Most of confirmed extra-galactic GeV-TeV sources are blazars. The bimodal feature of their broadband SEDs is generally represented with the leptonic models of the synchrotron radiation and the inverse Compton (IC) scattering process (e.g. \citealt{Ghisellini+etal+1996, Urry+etal+1999}). The seed photons for the IC process can be from the synchrotron radiation photon field (SSC, \citealt{Maraschi+etal+1992, Ghisellini+etal+1996, Urry+etal+1999, Zhang+etal+2012}) or the external photon field(EC; \citealt{Sikora+etal+1994, Sikora+etal+2009}). NLS1 galaxies were identified as a new class of GeV AGNs by the $Fermi$/LAT (\citealt{Abdo+etal+2009}). Their broadband SEDs can be also explained with synchrotron+IC leptonic jet model (\citealt{Abdo+etal+2009}), which are similar to that in FSRQs. In addition, their radiation physics and jet properties are also similar to that in FSRQs (\citealt{Sun+etal+2015}). By modeling the SEDs of 3C 279 (a typical FSRQ) and two NLS1s (PMN J0948+0022 and 1H 0323+342) in different stages, \cite{Zhu+etal+2016} found a universal correlation between Doppler factors ($\delta$) and peak luminosities ($L_{\rm c}$) of external Compton scattering bump.

Comparative studies the similarity between the jet radiations from GRBs and AGN outbursts have been presented. A uniform correlation between synchrotron luminosity ($L_{\rm syn}$) and $\delta$ in GRBs and blazars is found by \cite{Wu+etal+2011}. \cite{Wang+Wei+2011} showed similar spectral energy distribution between GRB and AGN jet. \cite{Nemmen+etal+2012} illustrated that AGN jets and GRB jets exhibit the same correlation between the jet power and the gamma-ray luminosity, (see also \citealt{Zhang+etal+2013a, Wang+etal+2014}). Such a correlation may be also extended to the jets in black hole X-ray binaries (BXBs) in hard/quiescent states and low-luminosity active galactic nuclei, making this correlation may exist for jets with luminosity spreading more than 20 orders of magnitude, i.e., from $10^{31}$ to  $10^{52}$ erg s$^{-1}$ (\citealt{Ma+etal+2014}). Further more, \cite{Zhu+etal+2019} found that the gamma-ray luminosity and power of outflows of short GRBs and pulsar wind nebulae follow that same relation, and the radiation efficiency is independent of the gamma-ray luminosity for various relativistic jet systems. They suggested that the acceleration and emission mechanisms or efficiencies may be similar in all relativistic outflows regardless of their central engines. \cite{Lyu+etal+2014} presented a unified picture for the radiation physics of relativistic jets in GRBs and blazars within the framework of the leptonic synchrotron radiation models. \cite{Zhang+etal+2017} proposed a potential fundamental plane for low-synchrotron-peak blazar and GRBs.

The distribution of blazars in the $\log L_{\rm syn, p}-E_{\rm syn}$ plane illustrate as a blazar sequence, i.e., high-luminosity FSRQs tend to have a low peak frequency and low-luminosity BL Lacs tend to have a high peak frequency. This sequence may be related to the different environments of emitting regions for different types of blazars (e.g. \citealt{Ghisellini+etal+1998}). However, positive correlation between $\log L_{\rm syn, p}$ and $E_{\rm syn}$ is observed for outbursts in individuals (\citealt{Massaro+etal+2008, Tramacere+etal+2009, Zhang+etal+2013a}). It is quite similar to that the $\log L_{\rm p}-E_{\rm p,z}$ relation in individual GRBs  (\citealt{Amati+etal+2002, Yonetoku+etal+2004, Liang+etal+2004, Ghirlanda+etal+2004, Lu+etal+2012}). Interestingly, \cite{Liang+etal+2015} found a tight correlation among the isotropic peak luminosity ($L_{\rm p}$), the peak energy ($E_{\rm p,z}$) of the $\nu f_\nu$ spectrum in the GRBs rest frame, and $\Gamma_0$ of GRBs jets. This $L_{\rm p}-E_{\rm p, z}-\Gamma_0$ relation is much tighter than the $\log L_{\rm p}-E_{\rm p,z}$ relation. This paper investigates whether or not GeV-TeV selected AGNs have a similar $L_{\rm syn, p}-E_{\rm syn, p}-\delta$ relation, and explore the possible implications for the physical origin of the $L_{\rm p}-E_{\rm p, z}-\Gamma_0$ relation of GRBs. We present our samples in Section 2. Our analysis results are present in Section 3. Discussion and conclusions are given in section 4.

\section{Sample and Data}

Our samples of GeV/TeV-selected FSRQs, BL Lacs, and NLS1s are taken from \cite{Zhang+etal+2012, Zhang+etal+2015} and \cite{Sun+etal+2015}. They presented systematical broadband SED fits to these AGNs with the single-zone lepton model. Viewing angle effect significantly influences the measurement of the Doppler factor of a radiating region. Since the jets in these AGNs are only middle relativistic and the viewing angle to the jet axis of blazars is through to be small, it is usually set $\delta=\Gamma$ in modeling the SEDs of blazars (e.g. \citealt{Zhang+etal+2012,Zhang+etal+2014}). Using the model parameters reported by \cite{Zhang+etal+2012, Zhang+etal+2015} and \cite{Sun+etal+2015}, we obtain the values of $\Gamma$, the peak luminosity ($L_{\rm syn}$) and the peak photon energy ($E_{\rm syn, p}$) in the source frame of the synchrotron emission. The bolometric luminosity ($L_{\rm bol}$) of these sources are also calculated with the SED fit results. Note that the model parameters for the BL Lacs are poorly constrained, and no error bars of the parameters are reported in \cite{Zhang+etal+2012}. Thus, no error is available for our data of BL Lacs. There are 18 FSRQs, 19 BL Lac objects, and 5 NLS1s are included in our AGN samples. Since these sources are variable, several SEDs in different outbursts are derived for some sources. The data of our AGN samples are reported in Tables ~\ref{Tab:FSRQs}-\ref{Tab:NLS1}. 
Thirty-four GRBs are included in our GRB sample. They are taken from \cite{Liang+etal+2015}. \cite{Liang+etal+2015} calculated the fireball initial Lorentz factors of these GRBs with the observed onset bump in the early optical afterglow lightcurves assuming that the onset bump is due to the deceleration of the fireballs by their ambient medium \cite{Sari+etal+1999}. The peak luminosity ($L_{\rm p}$) and the corresponding photon energy ($E_{\rm p}$) of these GRBs are derived from the fits to the observed spectra accumulated in 1-second peak time slice with the Band function or a power law with an exponential cutoff model. The bolometric luminosity in $1-10^4$ keV is calculated with the flux from spectral fits to the time-integrated spectra of these GRBs. The data are reported in Table~\ref{Tab:GRB}.

\begin{table}
\begin{center}
  	\caption{Data of Our FSRQs Sample$^{a}$}
	\label{Tab:FSRQs}
	\begin{tabular}{llllll}
    \hline\hline
  Name  & $z$ &  $\delta$  &  $\log(E_{\rm syn,z})$    &      $\log(L_{\rm syn,52})$ & $\log(L_{\rm bol,52})$ \\
          &  &   &  $(\rm keV)$  &  $(\rm erg \ s^{-1})$ & $(\rm erg \ s^{-1})$\\
  \hline
  {\bf FSRQs}&&&&\\
   \hline
  3C 279         & 0.536    &  12.0$\pm$0.5     &  $-4.29\pm0.15$   &   $-5.56\pm0.05$   & $-4.20\pm0.01$\\
  3C 273         & 0.158    &  7.4$\pm$0.9      &  $-3.97\pm0.20$   &   $-5.75\pm0.13$   & $-4.89\pm0.03$\\
  3C 454.3       & 0.859    & $17.6\pm0.6$      & $-4.13\pm0.10$    &   $-4.40\pm0.10$   & $ -2.94\pm0.01$\\
  PKS 1454-354   &  1.424   & $20.2\pm1.8$      & $-3.83\pm0.40$    &   $-4.65\pm0.4$    & $-3.08\pm0.04$\\
  PKS 0208-512   & 1.003    & $15.2\pm1.3$      & $-4.22\pm0.40$    &   $-5.13\pm0.40$   & $-3.66\pm0.02$\\
  PKS 0454-234   & 1.003    & $20.0\pm1.9$      & $-4.18\pm0.30$    & $-5.06\pm0.30$     & $-3.47\pm0.02$\\
  PKS 0727-11    & 1.589    & $20.6\pm1.2$      & $-3.97\pm0.20$    & $-4.82\pm0.15$     & $-3.09\pm0.03$\\
  PKS 0528+134   & 2.07     & $18.4\pm1.3$      & $-4.10\pm0.20$    & $-4.46\pm0.14$     & $-3.01\pm0.06$\\
  4C 66.2        & 0.657    & $12.2\pm1.2$      & $-4.21\pm0.48$    & $-5.59\pm0.15$     & $-4.23\pm0.02$\\
  4C 29.45       & 0.729    & $11.6\pm1.0$      & $-3.65\pm0.25$    & $-5.41\pm0.17$     & $-4.40\pm0.03$\\
  B2 1520+31     & 1.487    & $20.8\pm1.6$      & $-3.99\pm0.30$    & $-5.27\pm0.13$     & $-3.36\pm0.05$\\
  PKS 0420-01    & 0.916    & $12.8\pm0.7$      & $-3.66\pm0.30$    & $-5.23\pm0.13$     & $-3.99\pm0.02$\\
  1Jy 1308+326   & 0.997    & $12.6\pm0.9$      & $-4.13\pm0.35$    &$-5.76\pm0.20$      & $ -3.80\pm0.02$\\
  PKS 1510-089   & 0.36     & $11.0\pm0.5$      & $-4.30\pm0.06$    & $-6.14\pm0.05$     & $-4.54\pm0.03$\\
  4C 28.07       & 1.213    & $14.6\pm1.1$      & $-4.13\pm0.20$    & $-5.16\pm0.17$     & $-3.91\pm0.02$\\
  PMN 2345-1555  & 0.621    & $13.8\pm1.3$      & $-4.30\pm0.25$    & $-5.87\pm0.13$     & $-4.54\pm0.04$\\
  S3 2141+17     & 0.213    & $8.0\pm1.0$       & $-3.44\pm0.30$    & $-5.97\pm0.11$     & $-5.22\pm0.02$\\
  S4 0133+47     & 0.859    & $13.1\pm1.2$      & $-4.16\pm0.35$    &$-5.20\pm0.13$      & $-4.00\pm0.01$\\
  S4 0917+44     & 2.19     & $18.2\pm1.3$      & $-3.92\pm0.30$    & $-4.54\pm0.15$     & $-3.23\pm0.02$\\
  PKS 0227-369   & 2.115    & $17.8\pm1.0$      & $-3.79\pm0.30$    & $-4.69\pm0.13$     & $-3.20\pm0.03$\\
  PKS 0347-211   & 2.944    & $26.2\pm1.5$      & $-3.74\pm0.30$    & $-4.25\pm0.15$     & $-2.87\pm0.03$\\
  PKS 2325+093   &  1.843   & $17.6\pm17.6$     & $-3.33\pm0.30$    & $-3.90\pm0.13$     & $-3.16\pm0.03$\\
  PKS 1502+106   & 1.839    & $27.0\pm2.3$      & $-3.98\pm0.32$    & $-4.40\pm014$      & $-2.64\pm0.04$\\
\hline
	\end{tabular}
\end{center}
   $^{a}$ $\delta$ is the Doppler boosting factor, $E_{\rm syn,z}$ is the synchrotron peak photon energy in the source frame, $L_{\rm syn}$ and $L_{\rm bol}$ are the synchrotron peak luminosity and bolometric luminosity, respectively. They are derived from the SED fits with the single-zone leptonic model as reported in \cite{Zhang+etal+2015}.\\\\
\end{table}

\begin{table}
\begin{center}
  	\caption{Data of BL Lacs Sample$^{a}$}
	\label{Tab:BL Lacs}
	\begin{tabular}{llllll}
    \hline\hline
  Name  & $z$ &  $\delta$  &  $\log(E_{\rm syn,z})$    &      $\log(L_{\rm syn,52})$ & $\log(L_{\rm bol,52})$ \\
          &  &   &  $(\rm keV)$  &  $(\rm erg \ s^{-1})$ & $(\rm erg \ s^{-1})$\\
  \hline
 {\bf BL Lacs}&&&&\\
\hline
  Mkn 421$^{\rm L}$        &   0.031   &   29  &   $-0.295$   &   $-7.126$ & $-6.201$\\
  Mkn 501$^{\rm L}$	         &	0.034	&	14	&	$-0.914$	      &	$-7.709$ & $-6.745$\\
  Mkn 501$^{\rm H}$          &	0.034	&	15	&	$1.871$	      &	$-6.536$ & $-5.699$\\
  W Com$^{\rm L}$	         &	0.102	&	15	&	$-2.159$	&	$-7.090$ & $-6.000$\\
  W Com$^{\rm H}$            &	0.102	&	14	&	$-2.320$	&	$-6.995$ & $-5.553$\\
  BL Lacertae$^{\rm L}$ 	 &	0.069	&	19	&	$-3.304$	&	$-6.937$ & $-6.000$\\
  BL Lacertae$^{\rm H}$      &   0.069   &   20  &  $-2.979$    &   $-7.310$ & $-6.036$\\
  PKS 2005-489$^{\rm H}$     &   0.071   &   42  &	$-1.633$	&	$-6.659$ & $-5.620$\\
  1ES 1959+650$^{\rm L}$	 &	0.048	&	11	&	0.038       &   $-7.050$ & $-6.180$\\
  1ES 1959+650$^{\rm H}$     &	0.048	&	12	&	1.786	    &   $-6.741$ & $-6.678$\\
  1ES 2344+514$^{\rm L}$	 &	0.044	&	13	&	$-1.045$	&	$-7.898$ & $-7.036$\\
  PKS 2155-304$^{\rm L}$ 	 &	0.116	&	50	&	$-1.354$	&	$-6.137$ & $-5.208$\\
  PKS 2155-304$^{\rm H}$     &	0.116	&	26	&	$-1.041$	&	$-5.876$ & $-4.180$\\
  1ES 1101-232$^{\rm L}$	 &	0.186	&	12	&	$-0.223$	&	$-6.459$ & $-5.638$\\
  3C 66A	                 &	0.44	&	24	&	$-1.653$	&	$-5.396$ & $-3.921$\\
  PG 1553+113	 		&	0.3	    &	32	&	$-1.696$	&	$-5.386$ & $-4.509$\\
  1ES 1218+30.4			&	0.182	&	20	&	$-0.867$	&	$-6.612$ & $-5.638$\\
  1ES 1011+496			&	0.212	&	13	&	$-0.236$	&	$-5.799$ & $-4.959$\\
  PKS 1424+240			&	0.5	    &	33	&	$-1.319$	&	$-4.952$ & $-4.180$\\
  1ES 0806+524			&	0.138	&	12	&	$-1.496$    &	$-7.126$ & $-6.161$\\
  Mkn 180	    		&	0.045	&	6	&	$-1.092$	&	$-8.123$ & $-7.174$\\
  RGB J0152+017			&	0.08	&	5	&	$-0.236$	&	$-8.114$ & $-6.921$\\
  H1426+428	    		&	0.129	&	8.5	&	0.472	    &	$-7.264$ & $-5.699$\\
  PKS 0548-322			&	0.069	&	6	&	0.263	    &	$-7.692$ & $-6.796$\\
\hline
	\end{tabular}
\end{center}
   $^{a}$ $\delta$ is the Doppler boosting factor, $E_{\rm syn,z}$ is the synchrotron peak photon energy in the source frame, $L_{\rm syn}$ and $L_{\rm bol}$ are the synchrotron peak luminosity and bolometric luminosity, respectively. They are derived from the SED fits with the single-zone leptonic model as reported in \cite{Zhang+etal+2012}. Sources marked with ``H'' or ``L" indicate the high and low states as defined in \cite{Zhang+etal+2012}. \\\\
\end{table}

\begin{table}
\begin{center}
  	\caption{Data of NLS1 Sample$^{a}$}
	\label{Tab:NLS1}
	\begin{tabular}{llllll}
    \hline\hline
  Name  & $z$ &  $\delta$  &  $\log(E_{\rm syn,z})$    &      $\log(L_{\rm syn,52})$ & $\log(L_{\rm bol,52})$ \\
          &  &   &  $(\rm keV)$  &  $(\rm erg \ s^{-1})$ & $(\rm erg \ s^{-1})$\\
  \hline
 {\bf NLS1}&&&&\\
\hline

  1H 0323+342(1)     & 0.0629 &	$2.8\pm0.6$    &  $-3.96\pm0.40$  &  $-7.32\pm0.30$ & $-6.30\pm0.01$\\
  1H 0323+342(2)     & 0.0629 &	$3.6\pm1.3$    &  $-4.43\pm0.45$  &  $-7.52\pm0.50$ & $-6.52\pm0.02$\\
  1H 0323+342(3)     & 0.0629 &	$4.9\pm0.8$    &  $-4.62\pm0.40$  &  $-7.29\pm0.40$ & $-6.37\pm0.01$\\
  1H 0323+342(4)     & 0.0629 &	$4.5\pm0.6$	   &  $-4.80\pm0.40$  &  $-7.34\pm0.40$ & $-6.26\pm0.01$\\
  1H 0323+342(5)     & 0.0629 & $6.2\pm0.6$	   &  $-4.76\pm0.15$  &  $-6.98\pm0.15$ & $-5.95\pm0.01$\\
  PMN J0948+0022(1)  & 0.5846 & $11.1\pm1.4$   &  $-4.38\pm0.42$  &  $-5.60\pm0.25$ & $-4.60\pm0.02$\\
  PMN J0948+0022(2)  & 0.5846 &	$10.8\pm1.3$   &  $-4.20\pm0.25$  &  $-5.31\pm0.24$ & $-4.60\pm0.01$\\
  PMN J0948+0022(3)  & 0.5846 &	$8.6\pm1.3$	   &  $-4.22\pm0.40$  &  $-5.68\pm0.30$ & $-4.79\pm0.01$\\
  PMN J0948+0022(4)  & 0.5846 &	$11.1\pm1$     &  $-4.48\pm0.32$  &  $-5.38\pm0.25$ & $-4.65\pm0.02$\\
  PMN J0948+0022(5)  & 0.5846 &	$11.6\pm0.8$   &  $-4.45\pm0.25$  &  $-5.42\pm0.15$ & $-4.43\pm0.02$\\
  PMN J0948+0022(6)  & 0.5846 & $9.5\pm0.5$	   &  $-4.54\pm0.17$  &  $-5.95\pm0.13$ & $-4.67\pm0.02$\\
  PMN J0948+0022(7)  & 0.5846 &	$13.5\pm1.1$   &  $-4.89\pm0.20$  &  $-5.32\pm0.20$ & $-3.76\pm0.02$\\
  PMN J0948+0022(8)  & 0.5846 &	$13.7\pm1.8$   &  $-5.08\pm0.45$  &  $-5.28\pm0.34$ & $-3.92\pm0.02$\\
  PMN J0948+0022(9)  & 0.5846 &	$11.4\pm2.2$   &  $-4.08\pm0.40$  &  $-5.37\pm0.35$ & $-4.30\pm0.01$\\
  SBS 0846+513       & 0.5835 &	$7.4\pm0.8$    &  $-4.51\pm0.15$  &  $-6.67\pm0.09$ & $-5.15\pm0.03$\\
  PKS 1502+036       & 0.409  & $9.5\pm0.8$    &  $-4.29\pm0.20$  &  $-6.47\pm0.15$ & $-5.39\pm0.08$\\
  PKS 2004-447       & 0.24   & $6.4\pm0.5$    &  $-4.29\pm0.15$  &  $-6.97\pm0.10$ & $-6.18\pm0.02$\\

\hline
	\end{tabular}
\end{center}
   $^{a}$ $\delta$ is the Doppler boosting factor, $E_{\rm syn,z}$ is the synchrotron peak photon energy in the source frame, $L_{\rm syn}$ and $L_{\rm bol}$ are the synchrotron peak luminosity and bolometric luminosity, respectively. They are derived from the SED fits with the single-zone leptonic model as reported in \cite{Sun+etal+2015}. Different flux states of two Narrow Line Seyfert 1 galaxies, 1H 0323+342 and PMN J0948+0022, are also reported in \cite{Sun+etal+2015}.\\\\
\end{table}

\begin{table}
\begin{center}
	\caption{Data of Our GRB Sample Taken from Liang et al. (2015)$^{a}$}
	\label{Tab:GRB}
	\begin{tabular}{lllllll}
    \hline\hline
  GRB  & $z$ & $T_{\rm 90}$ & $\Gamma_0$  &  $\rm log(E_{\rm p,z})$      &      $\log(L_{\rm p,52})$ & $\log(L_{\rm bol,52})$  \\
          &  & (s)  & &  $(\rm keV)$  &  $(\rm erg \ s^{-1})$  & $(\rm erg \ s^{-1})$\\
 \hline
  990123  & 1.6 &	$63.3\pm0.26$    &  $600\pm80$  &  $3.13\pm0.02$ &   $1.44\pm0.02$  &$0.75\pm0.01$ \\
  090924  & 0.544 &	$48\pm3$         &  $300\pm79$  &  $2.44\pm0.01$ &   $0.32\pm0.03$  &$-1.04\pm0.03$ \\
  080810  &3.35   &  $106\pm5$        &  $409\pm34$  &  $3.13\pm0.10$ &   $0.98\pm0.04$  &$-0.42\pm0.03$ \\
  060605  &3.78   & $15\pm2$   &$197\pm30$ &$2.69\pm0.22$  & $-0.02\pm0.07$ & $-0.72\pm0.09$\\
  050820A &2.615  & $50\pm5$  & $282\pm29$ &$2.95^{+0.22}_{-0.12}$ & $0.51^{+0.04}_{-0.06}$ & $-0.50\pm0.05$\\
  060607A &3.082  &  $100\pm5$ & $296\pm28$ & $2.76\pm0.15$ & $0.30\pm0.06$ & $-1.05\pm0.34$\\
  060418  &1.489  & $52\pm1$  & $263\pm23$  & $2.76^{+0.23}_{-0.06}$ & $0.28\pm0.03$ & $-0.56\pm0.01$\\
  070208  &1.165  & $48\pm2$  & $115\pm23$  & $1.82^{+1.18}_{-0.22}$ & $-1.03\pm0.05$ & $-2.23\pm0.34$\\
  081203A &2.1    & $294\pm71$ & $219\pm21$ & $3.19\pm0.21$  & $0.45\pm0.03$ & $-0.92\pm0.11$\\
  070419A &0.97   & $116\pm6$ &  $91\pm11$  & $1.43^{+0.26}_{-0.31}$ & $-2.01\pm0.04$ &$-1.82\pm0.43$\\
  060904B &0.703  & $192\pm5$ & $108\pm10$  & $2.13\pm0.13$ & $-1.13\pm0.08$ & $-2.72\pm0.09$\\
  080710  &0.845  & $120\pm17$ & $53\pm8$   & $2.48^{+0.72}_{-0.29}$ & $-1.10\pm0.04$ & $-2.18\pm0.44$\\
  080319C &1.95   & $34\pm9$   & $228\pm5$  & $3.24\pm0.13$ & $0.98\pm0.01$ & $-0.36\pm0.12$\\
  071010B &0.947  & $35.7\pm0.5$ & $209\pm4$ & $2.01\pm0.05$ & $-0.26\pm0.02$ & $-1.14\pm0.03$\\
  070110  &2.352  & $85\pm5$  & $127\pm4$ &  $2.57\pm0.20$  & $-0.35\pm0.07$ & $-1.19\pm0.12$\\
  060210  &3.91   & $46\pm10$ & $264\pm4$ &  $2.86^{+1.17}_{-0.10}$ & $0.87^{+0.12}_{-0.08}$ & $0.08\pm0.17$\\
  061007  &1.261  & $75\pm5$  & $436\pm3$ &  $2.96\pm0.02$ & $1.16^{+0.06}_{-0.07}$ & $0.10\pm0.04$\\
  061121  &1.314  & $81\pm5$  &  $175\pm2$ & $3.11\pm0.05$ & $1.15\pm0.01$ & $-0.49\pm0.06$\\
  090812  &2.452  & $66.7\pm14.7$ & $501\pm46$ & $3.30^{+0.19}_{-0.12}$ & $1.00^{+0.04}_{-0.06}$ & $-0.20\pm0.11$\\
  060218  &0.0331 & $100\pm10$ & $2.3\pm0.3$ & $0.71\pm0.03$ & $-5.37\pm0.16$ & $-5.03\pm0.06$\\
  100621A &0.542  & $63.6\pm1.7$ & $52.0\pm4.8$ & $2.16\pm0.07$ & $-0.50\pm0.03$ & $-1.16\pm0.05$\\
  050922C &2.198  & $5\pm1$  & $274\pm25$ & $2.80^{+0.14}_{-0.08}$ & $0.82^{+0.02}_{-0.03}$ & $0.04\pm0.09$\\
  091029  &2.752  & $39.2\pm5.0$ & $221\pm20$ & $2.36\pm0.13$ & $0.24\pm0.03$ & $0.62\pm0.07$\\
  071112C &0.822  & $15\pm2$ & $244\pm22$ & $2.63^{+0.14}_{-0.09}$ & $0.02\pm0.04$ & $-0.84\pm0.13$\\
  080129  &4.394  & $48\pm10$ & $65\pm6$  & $3.13^{+0.85}_{-0.26}$ & $0.43\pm0.04$ & $-0.84\pm0.13$\\
  081109A &0.98   & $190\pm60$ & $68\pm7$ & $2.32^{+0.63}_{-0.10}$ & $-0.71\pm0.06$ & $-1.67\pm0.25$\\
  081008  &1.967  & $185.5\pm40.3$ & $250\pm23$ & $2.43^{+0.55}_{-0.10}$ & $-0.26\pm0.01$ & $-1.44\pm0.14$\\
  091024  &1.092  & $109.8\pm16.7$  &  $69\pm6$  & $2.90\pm0.13$ & $0.21\pm0.10$ & $-0.59\pm0.08$\\
  090102  &1.547  & $27.0\pm2.2$  & $61\pm6$ & $3.06^{+0.07}_{-0.06}$ & $0.77\pm0.06$ & $-0.10\pm0.05$\\
  110205A &2.22   & $257\pm25$ & $177\pm16$  & $2.85\pm0.15$  & $0.40\pm0.06$ & $-0.66\pm0.06$\\
  121217A &3.1    & $778\pm16$ & $247\pm23$  &  $2.88\pm0.13$ & $0.55\pm0.07$ & $-1.10\pm0.10$\\
  100728B &2.106  & $12.1\pm2.4$ & $373\pm34$ & $2.61\pm0.03$ & $0.27\pm0.03$ & $-0.61\pm0.10$\\
  110213A &1.46   & $48\pm16$ & $223\pm21$ & $2.38\pm0.02$ & $0.32\pm0.01$ & $-0.88\pm0.15$\\
  100906A &1.727  & $114.4\pm1.6$ & $369\pm34$ & $2.20\pm0.04$ & $0.39\pm0.02$ & $-0.53\pm0.04$\\
\hline
	\end{tabular}
\end{center}
$^{a}$ $T_{90}$ is the GRB duration, $\Gamma_0$ is the initial Lorentz factor of the GRB firballs, $E_{\rm p, z}$ is the peak photon energy of the GRBs derived from the fits with the Band function (\citealt{Band+etal+1993}) in the burst frame, $L_{\rm p}$ and $L_{\rm bol}$ are the luminosity at the 1-second peak time slice and the time-integrated luminosity in the burst duration, respectively.
\end{table}

\section{Correlation Analysis Results}

\begin{figure}[htbp]
\centering
\includegraphics[width=0.45\textwidth, angle=0]{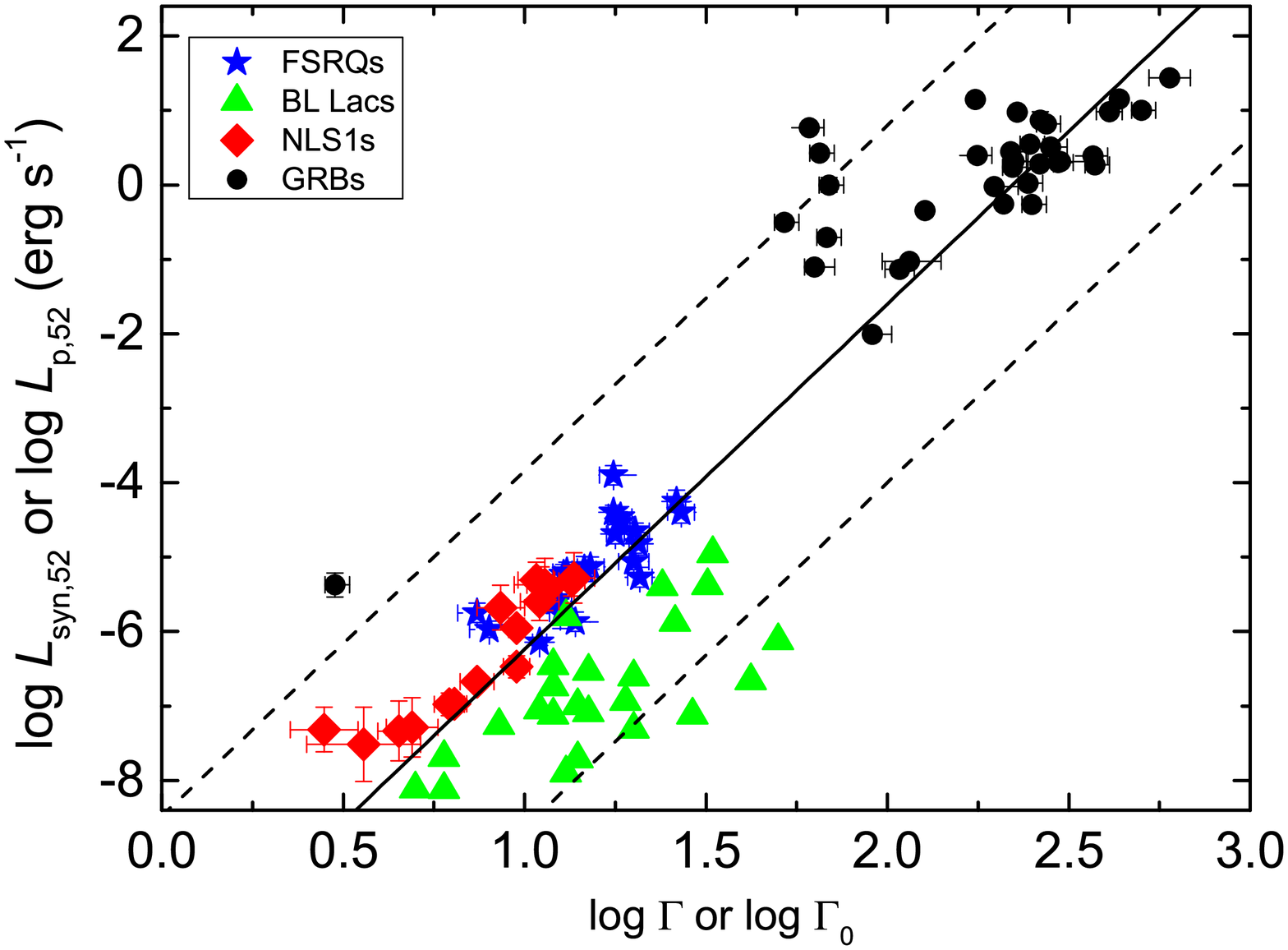}
\includegraphics[width=0.45\textwidth, angle=0]{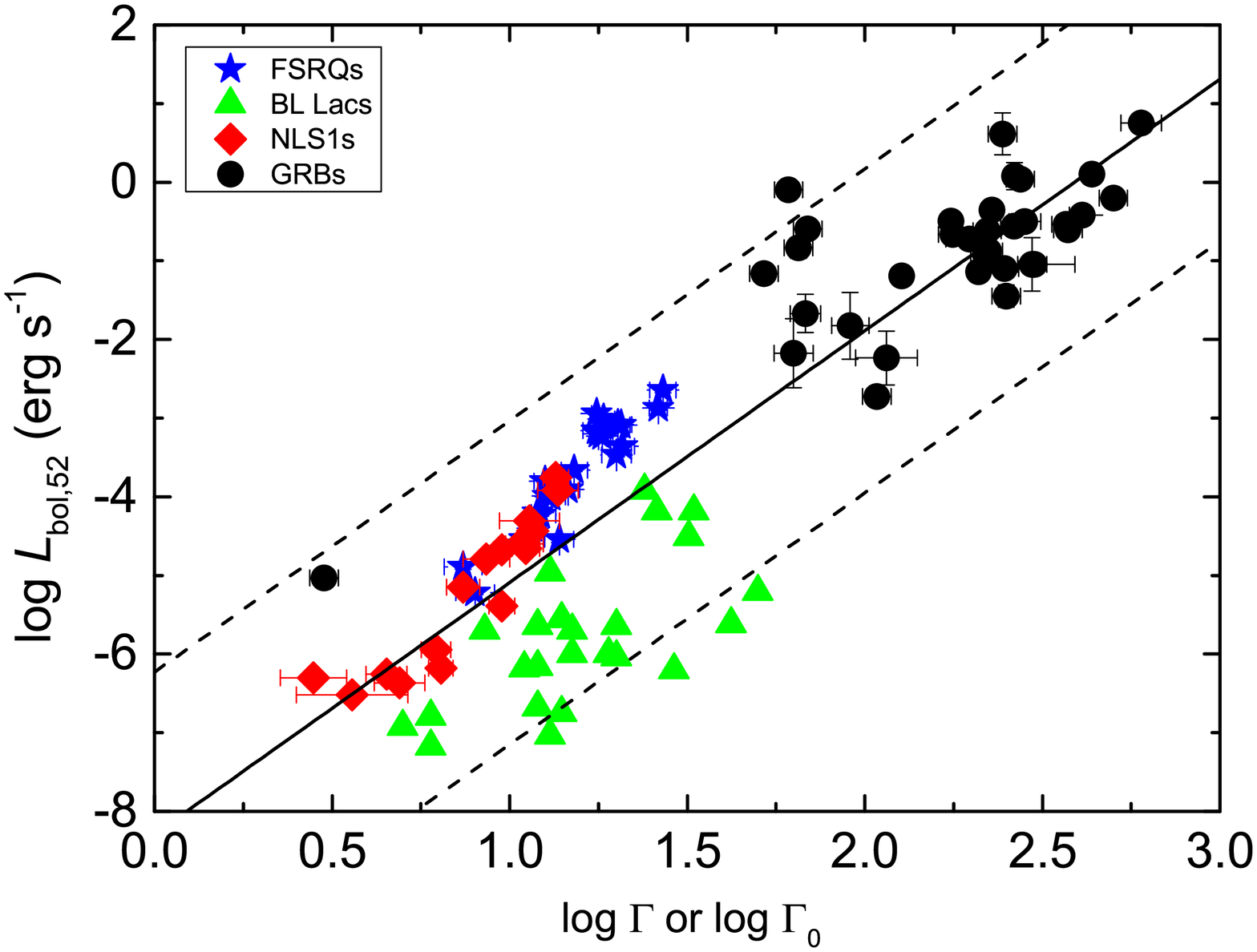}
\caption{Synchrotron peak luminosity and bolometric luminosity in the observed frames as a function of the jet Lorentz factor for the AGNs in our samples. GRBs in our sample are illustrated accordingly with their 1-second peak time luminosity ($L_{\rm p}$), time-integrated luminosity in the energy band of $1-10^{4}$ keV. Lines are the best fit and the 2 $\sigma$ dispersion derived from the Spearman linear correlation analysis for both the AGNs and GRBs.}
\label{Lsyn-Gamma}
\end{figure}

We make Spearman pair correlation analysis between the luminosity and Doppler boosting factor for each sub-groups of the ANGs and for the entire samples of the AGNs and GRBs. Our results are reported in Table \ref{Tab:pair}. It is found that both $L_{\rm syn}$ (or $L_{\rm p}$) and $L_{\rm bol}$ depend on $\Gamma$ (or $\Gamma_0$) with a power-law index ranging from $2.27$ to $4.58$ for different sub-classes of AGNs. We make correlation analysis for the entire AGN and GRB samples, as shown in Figure~\ref{Lsyn-Gamma}, the BL Lacs are separated from the FSRQs and NLS1 galaxies, and BL Lacs tend to be dimmer than the FSRQs and NLS1 galaxies with the same $\Gamma$. Both GRBs and AGNs shape a clear sequence in the $\log L_{\rm syn}-\log \Gamma$ and $\log L_{\rm bol}-\log \Gamma$ planes. Our correlation analysis yields $\log L_{\rm syn} \propto \Gamma^{4.64\pm 0.20}$, and $\log L_{\rm bol} \propto \Gamma^{3.20\pm 0.17}$ (see also \citealt{Wu+etal+2011}). However, this relation has very large dispersion ($\Delta=1.20$). Our Spearman correlation analysis between $E_{\rm syn}$ and $\delta$ does not reveal any statistical correlation with a chance probability $p<10^{-4}$ between the two quantities in each sub-class of the AGNs.

\begin{figure}[htbp]
\centering
\includegraphics[width=0.45\textwidth, angle=0]{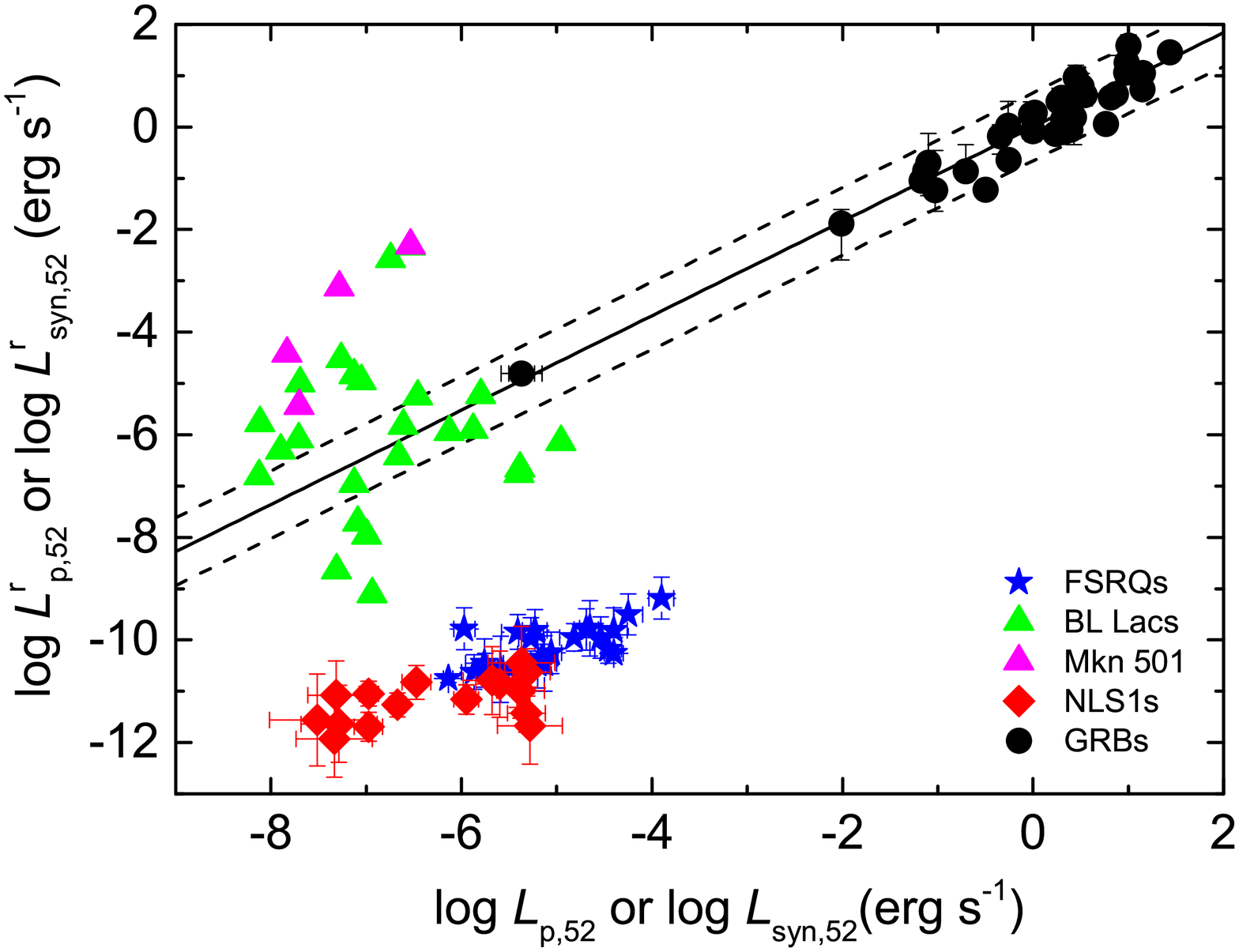}
\includegraphics[width=0.45\textwidth, angle=0]{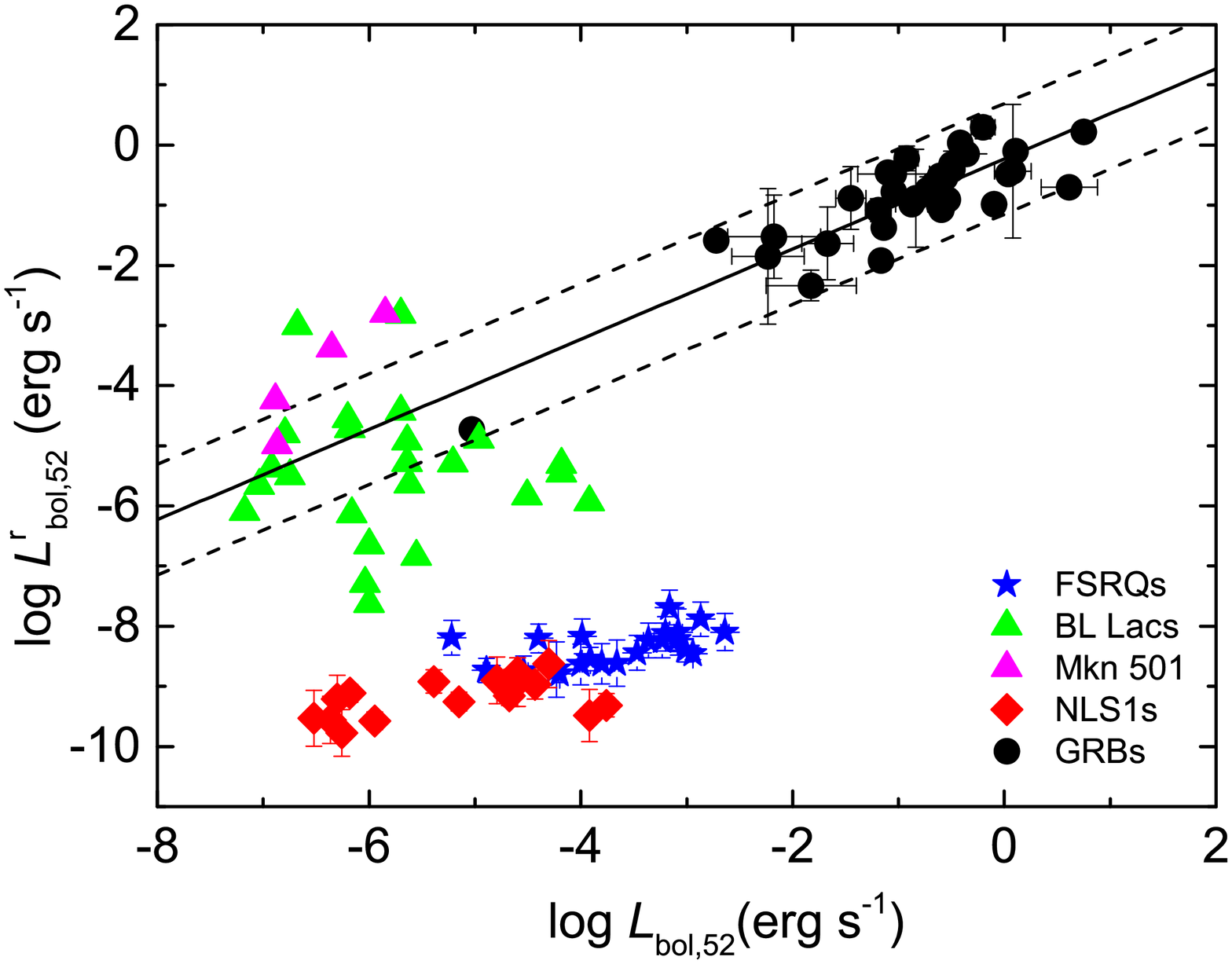}
\caption{Examination of whether or not the jet radiations of the AGNs share the same $L-E_{\rm p}-\Gamma_0$ relation as that derived from GRBs, in which  $L^{\rm r}_{\rm syn}$ and $L^{\rm r}_{\rm bol}$ are calculated with the relations of $L^{\rm r}_{\rm p}(E_{\rm p}, \Gamma_0)$ or  $L^{\rm r}_{\rm bol}(E_{\rm p}, \Gamma_0)$ derived from the GRB sample. The best linear fit line together with their 2$\sigma$ dispersion regions of the relations are shown with solid and dashed lines, respectively (e.g. \citealt{Liang+etal+2015}). The pink triangles are the data for Mkn 501 in different outbursts taken from \cite{Zhang+etal+2013b}.}
\label{L-E-Gamma-GRB}
\end{figure}

Physically, the observed luminosity and photon energy are boosted by the jet bulk Doppler effect. As shown in \cite{Liang+etal+2015}, by incorporating the Doppler boosting factor the derived $L_{\rm p}-E_{\rm p}-\Gamma_{0}$ relation is much tighter than the $L_{\rm p}-E_{\rm p}$ relation. We first examine whether the synchrotron radiations of the selected AGNs follow the $L_{\rm p}-E_{\rm p}-\Gamma_{0}$ relation of GRBs, i.e., $L_{\rm p}\propto E_{\rm p,z}^{1.34\pm0.14}\Gamma_0^{1.32\pm0.19}$ (\citealt{Liang+etal+2015}). We calculate the synchrotron peak luminosity ($L^{\rm r}_{\rm syn}$) with this relation for the AGNs by using their $E_{\rm syn, p}$ and $\Gamma$ values. Figure \ref{L-E-Gamma-GRB} shows $L^{\rm r}_{\rm syn}$ as a function of the observed $L_{\rm syn}$. It is found that the BL Lacs are in the low luminosity end of this relation with a very large scatter, and the derived  $L^{\rm r}_{\rm syn}$ of FSRQs and NLS1 galaxies are 4-5 orders of magnitude lower than the $L_{\rm p}-E_{\rm p}-\Gamma_{0}$ relation of GRBs. The FSRQs and NLS1 seem to follow another tight relation which is different from that of GRBs.

\begin{figure}[htbp]
\centering
\includegraphics[width=0.45\textwidth, angle=0]{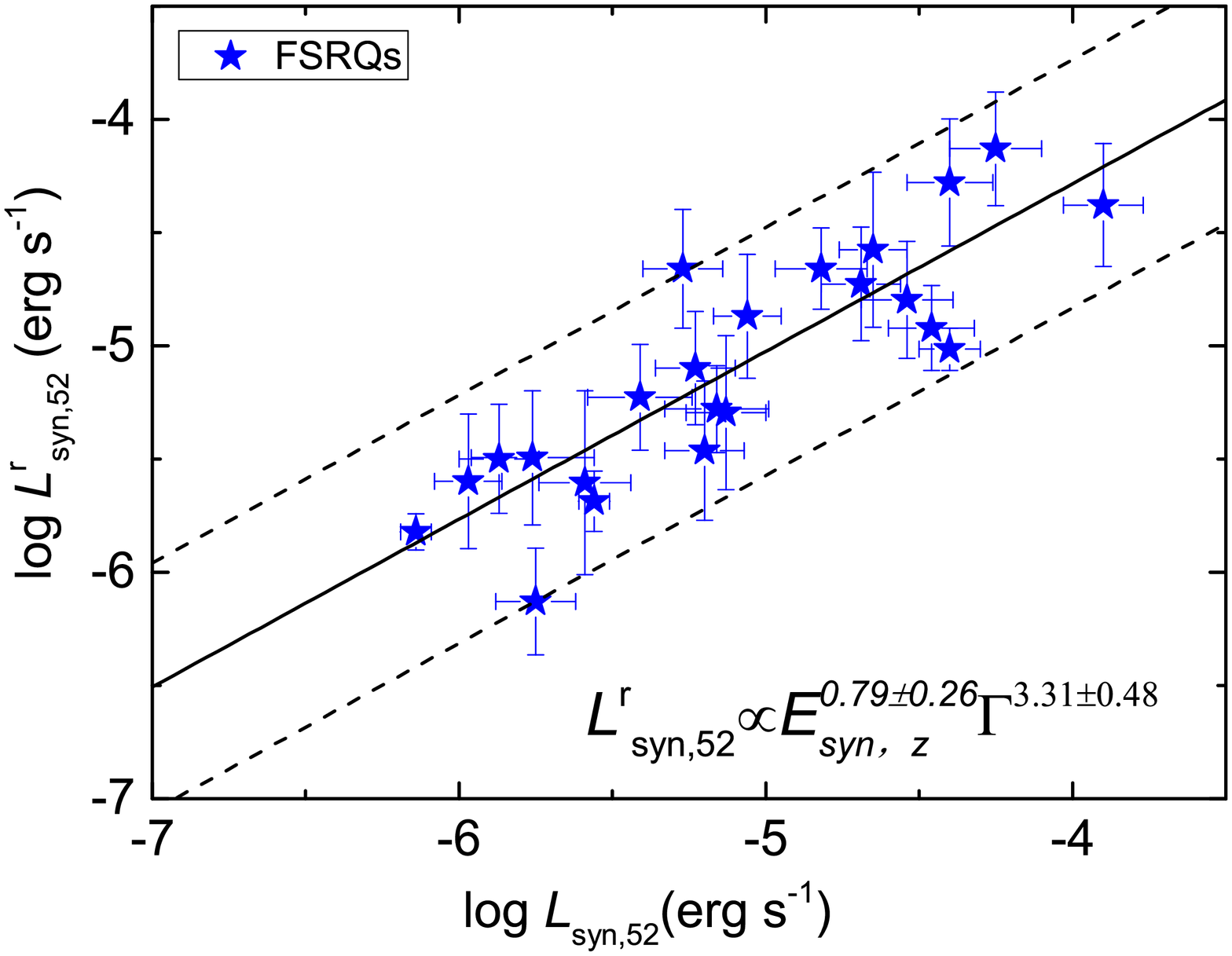}
\includegraphics[width=0.45\textwidth, angle=0]{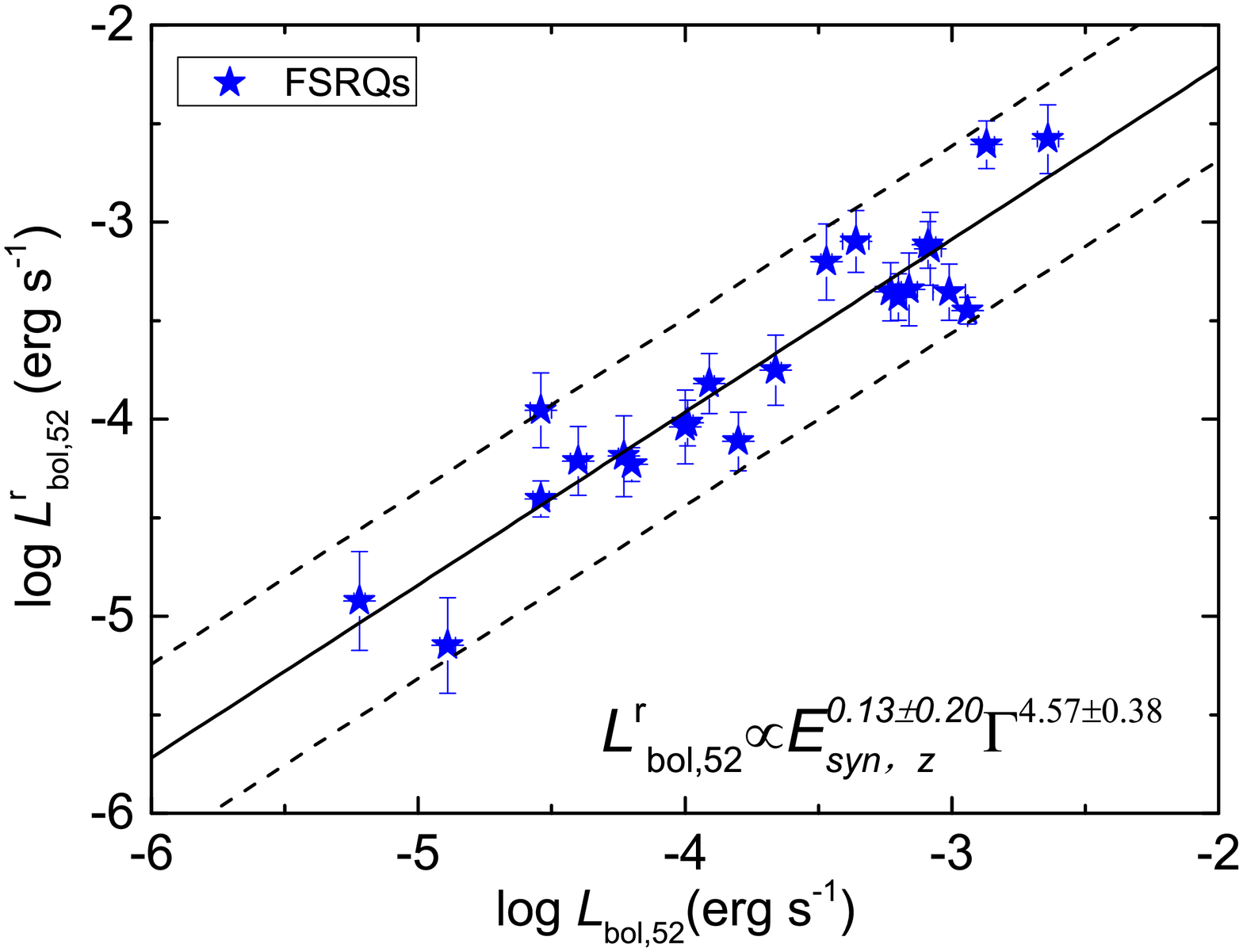}
\includegraphics[width=0.45\textwidth, angle=0]{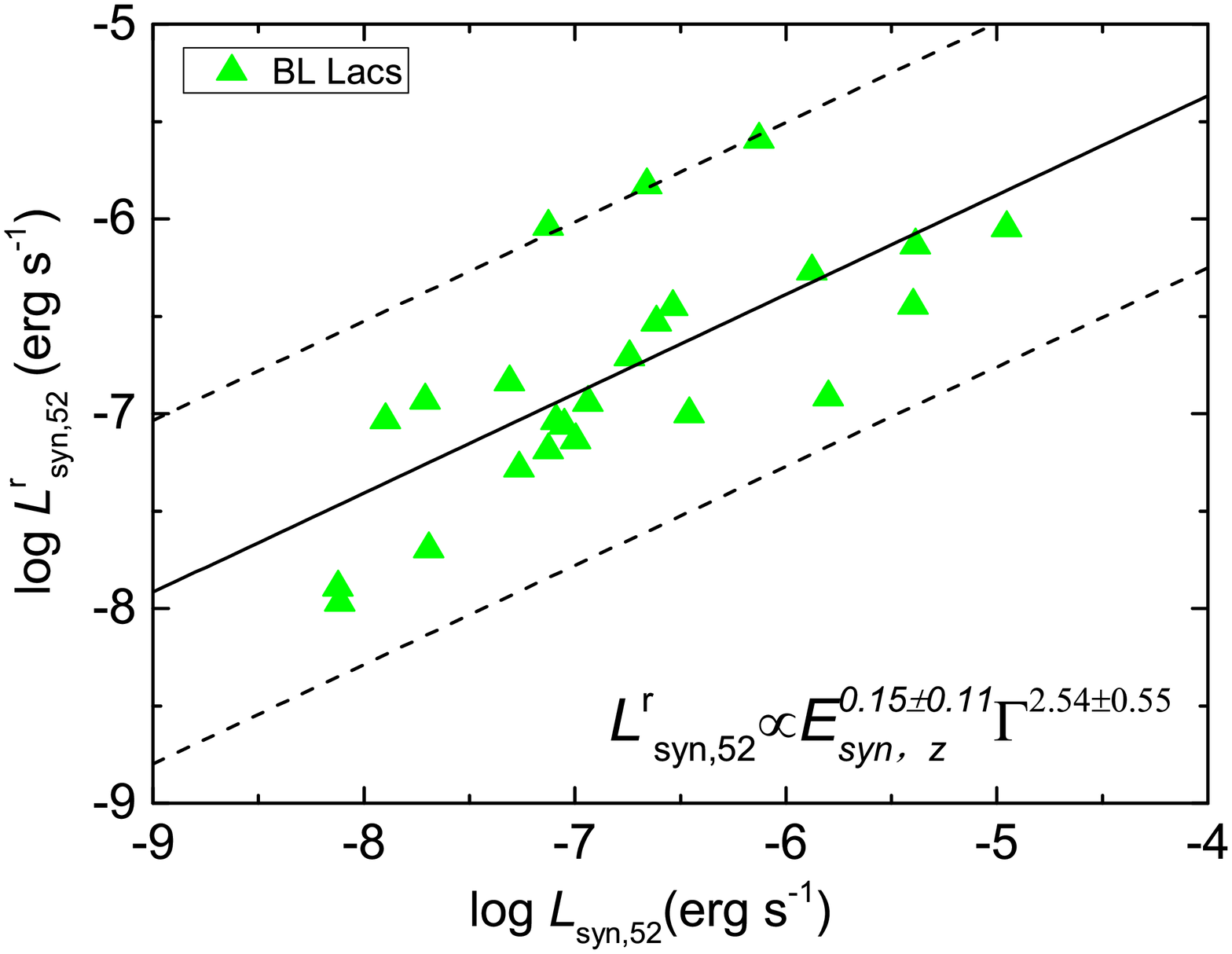}
\includegraphics[width=0.45\textwidth, angle=0]{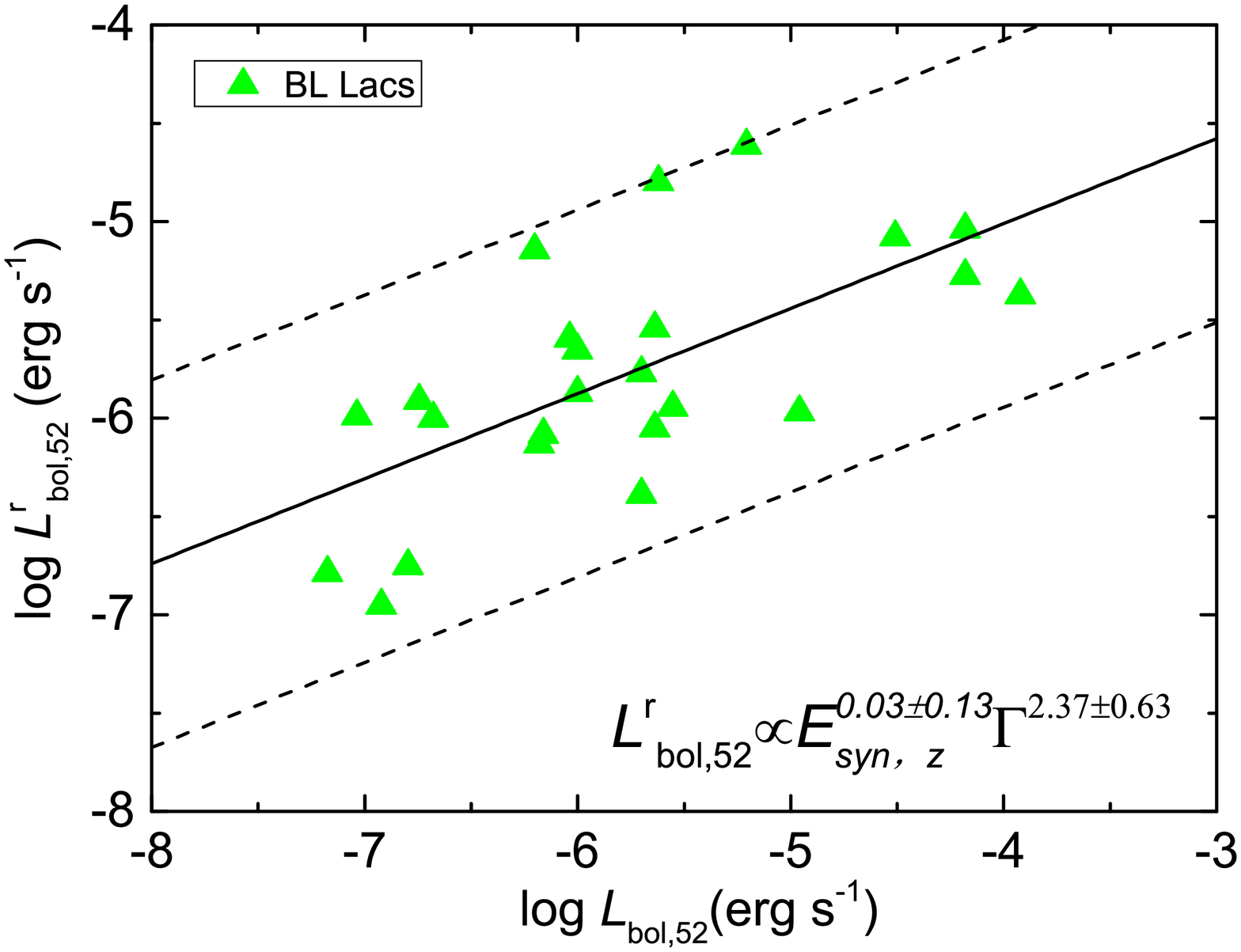}
\includegraphics[width=0.45\textwidth, angle=0]{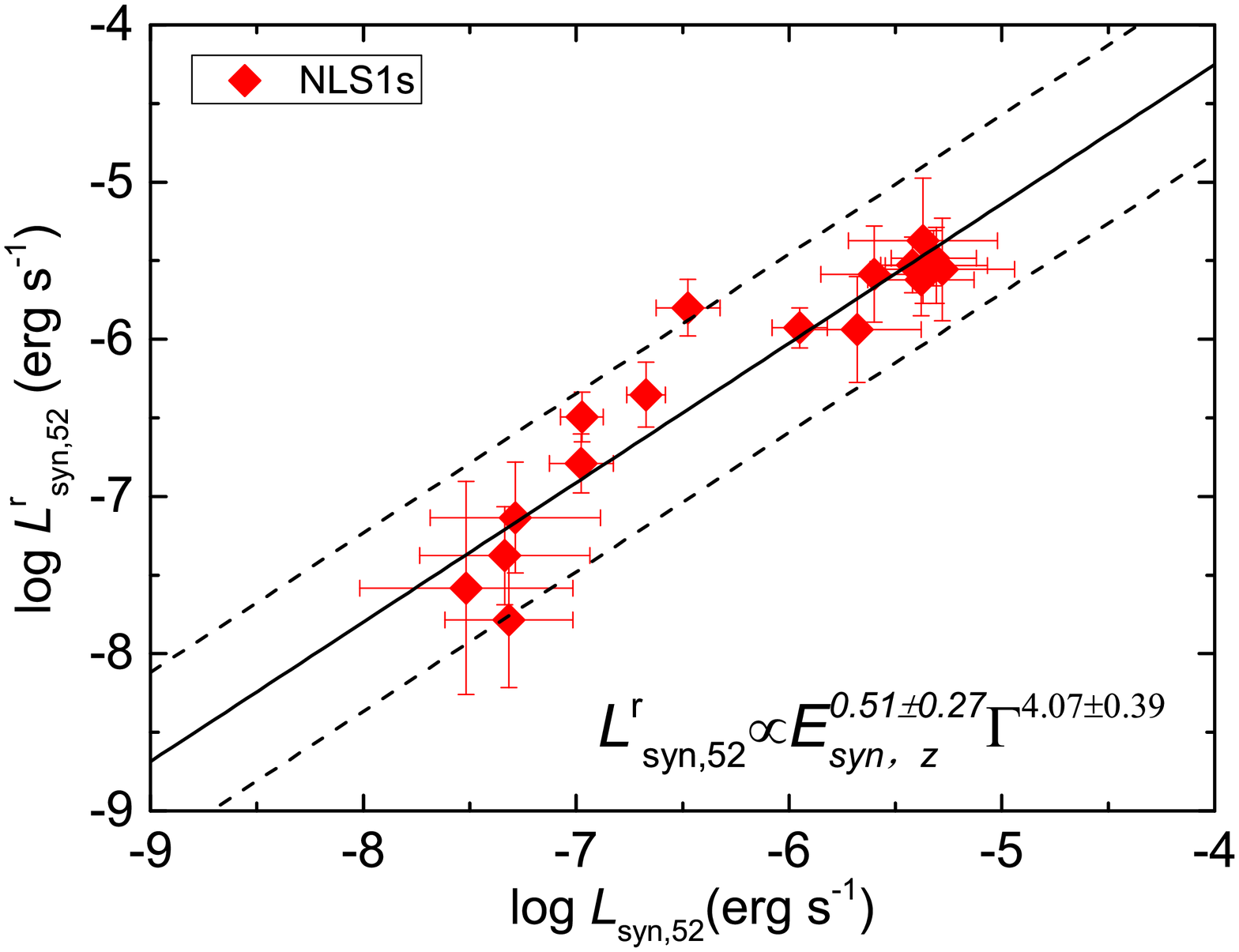}
\includegraphics[width=0.45\textwidth, angle=0]{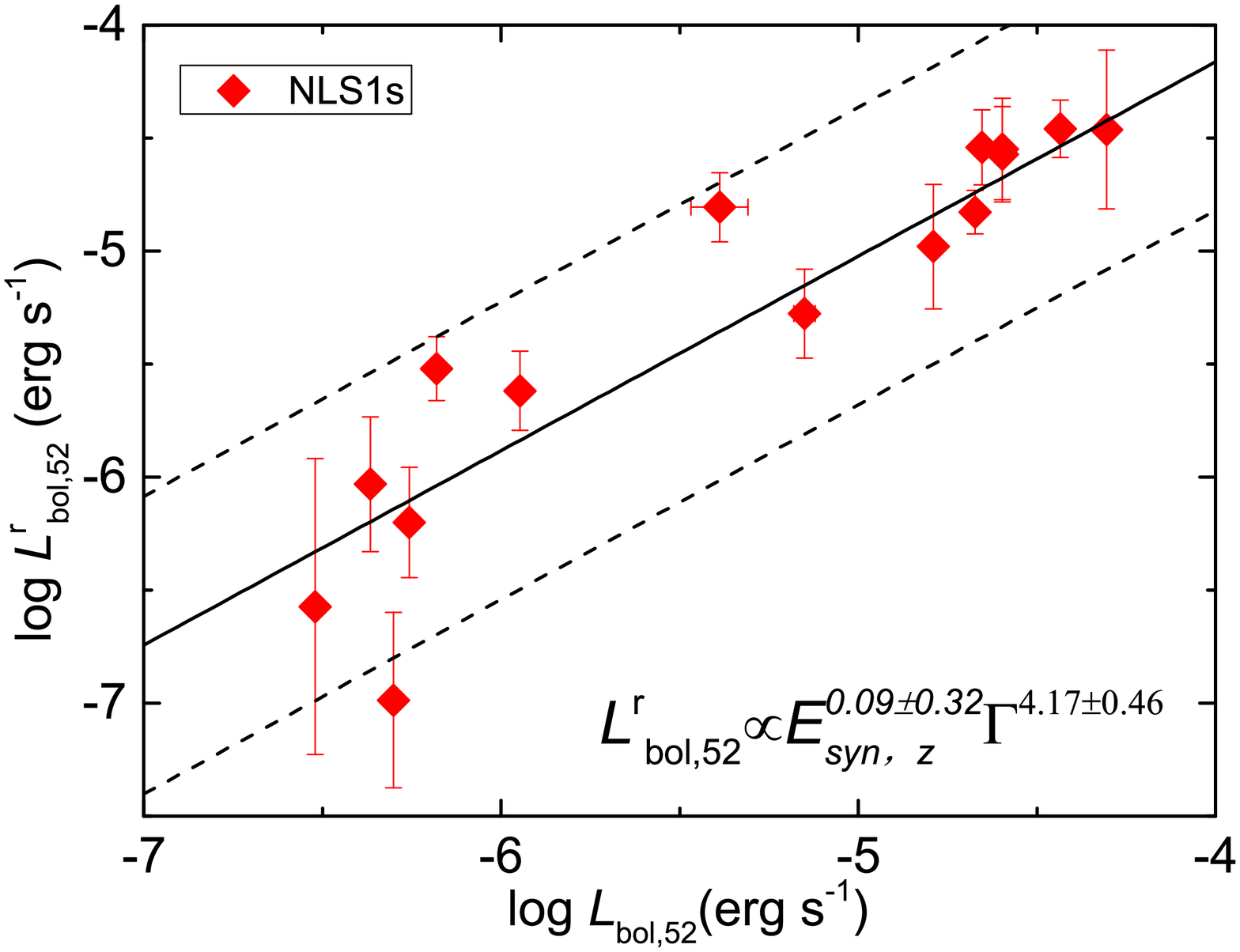}
\caption{Pair correlations of $\log L^{\rm r}_{\rm syn}-\log L_{\rm syn}$ and $\log L^{\rm r}_{\rm bol}-\log L_{\rm bol}$ planes, where $L^{\rm r}_{\rm syn}$ and $L^{\rm r}_{\rm bol}$ are calculated with the relations of $L^{\rm r}_{\rm syn}(E_{\rm syn, p}, \Gamma)$ or  $L^{\rm r}_{\rm bol}(E_{\rm syn, p}, \Gamma)$ derived from each subclasses of AGN samples, as marked in each panels. The best linear fit line together with their 2$\sigma$ dispersion regions of the relations are shown with solid and dashed lines, respectively.}
\label{L-E-Gamma-AGN}
\end{figure}

\begin{figure}[htbp]
\centering
\includegraphics[width=0.45\textwidth, angle=0]{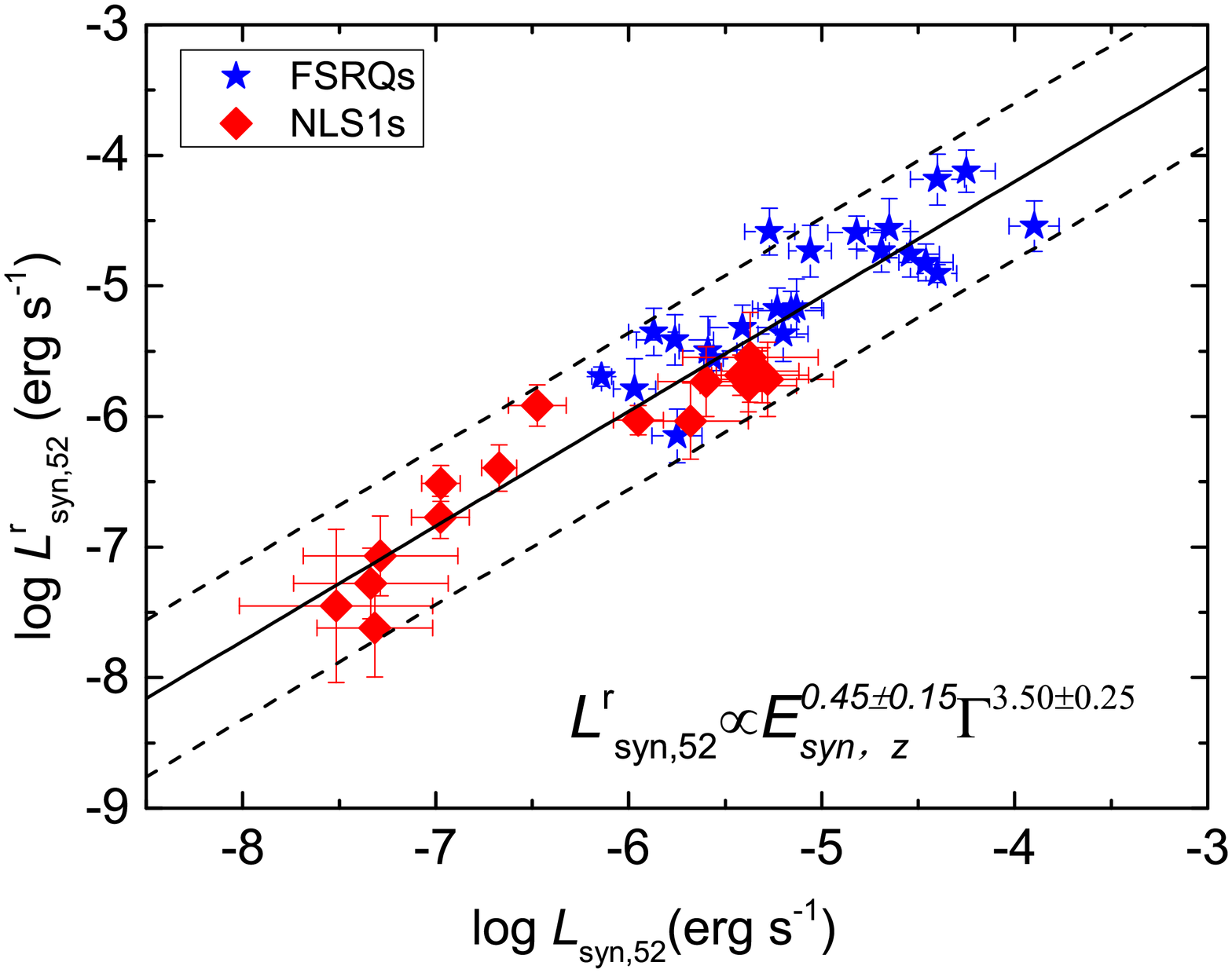}
\includegraphics[width=0.45\textwidth, angle=0]{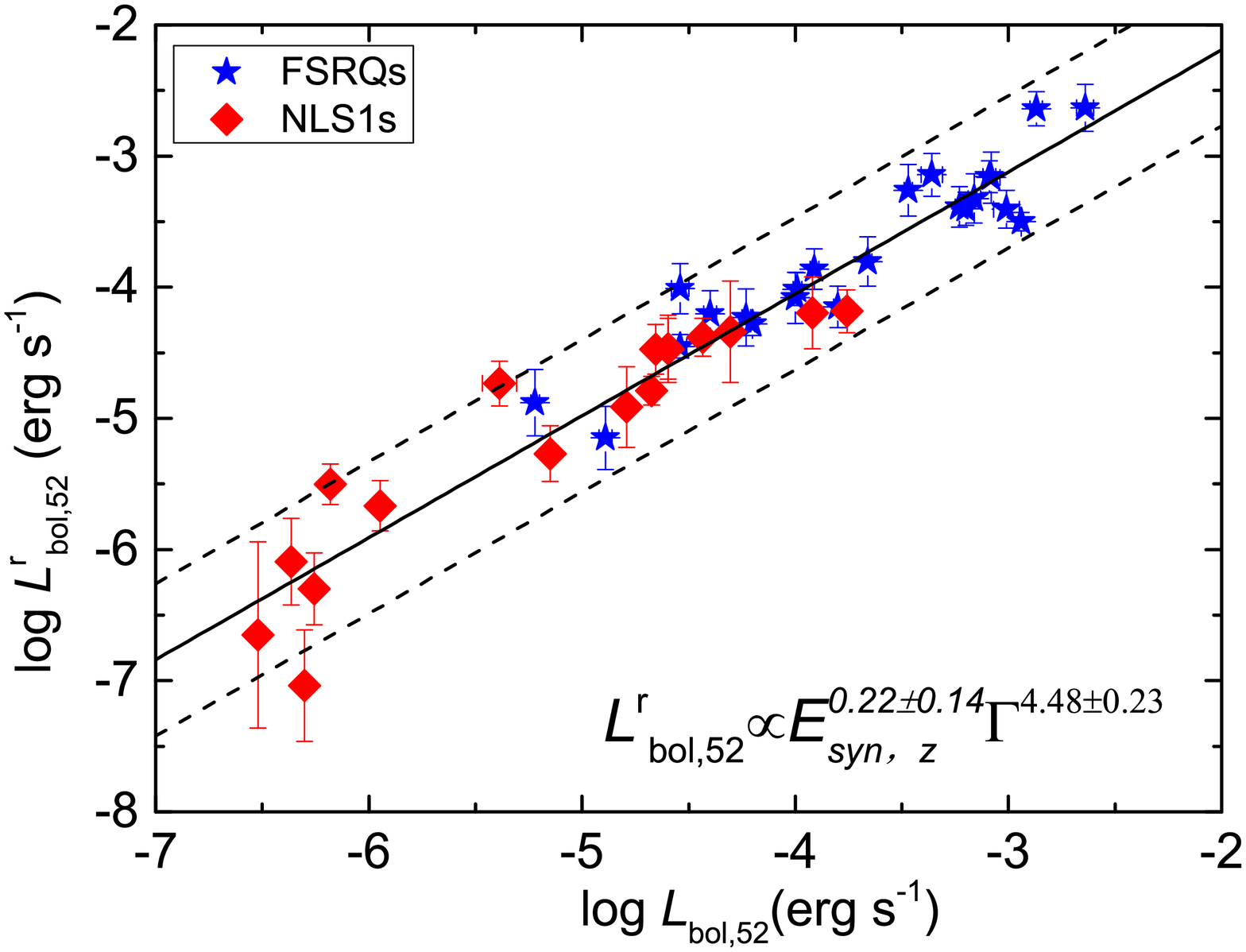}
\caption{The same as Figure~\ref{L-E-Gamma-GRB} but for the combined sample of the FSRQs and Narrow line Seyfert 1 galaxies.}
\label{L-E-Gamma-FSRQ-NLS1}
\end{figure}

We explore $L-E_{\rm syn, p}-\Gamma$ relation for each sub-classes of the AGNs, using the stepwise regression analysis method. Our model is $\log L^{r} (\log E_{\rm syn, p},\log \Gamma)=a+b\log E_{\rm syn, p}+c\log \Gamma$. Our results are reported in Table \ref{Tab:triple} and shown in Figure \ref{L-E-Gamma-AGN}. We do not find a $L-E_{\rm syn, p}-\Gamma$ relation with $p_{\rm F}<10^{-4}$ for the BL Lacs, where $p_{\rm F}$ is the probability of the F-test for our regression analysis. Similar $L_{\rm syn}-E_{\rm syn, p}-\Gamma$ and $L_{\rm bol}-E_{\rm syn, p}-\Gamma$ relations are found for the FSRQs and NLS1 galaxies. Our regression analysis for the combined sample of the FSRQs and NLS1 galaxies yields $\log L_{\rm syn,52}=(-7.40\pm0.77)+(0.45\pm0.15)\log E_{\rm syn,z}/ \rm keV+(3.50\pm0.25)\log \Gamma$ and $\log L_{\rm bol,52}=(-8.16\pm0.71)+(0.22\pm0.14)\log E_{\rm syn,z}/ \rm keV+(4.48\pm0.23)\log \Gamma$, as shown in Figure~\ref{L-E-Gamma-FSRQ-NLS1}. One can find that the dispersion of the three parameter relations are significantly tighter than the $L-\Gamma$ relations.

\begin{table}
%\tiny
\begin{center}
	\caption{Results of our Spearman linear correlation analysis for the AGNs and GRBs in our samples. $r$ and $p$ are the linear correlation coefficient and chance probability, and $\Delta$ is the $1\sigma$ dispersion of the pair correlation.}
	\label{Tab:pair}
	\begin{tabular}{llllll}
    \hline\hline
     relations  &  Source &  $\rm Expressions$ &   $r$  &  $p$      &      $\Delta$ \\
     \hline

     $L_{\rm syn}(\Gamma)$ & FSRQs & $L_{\rm syn,52}=10^{(-9.04\pm0.68)}\Gamma^{(3.33\pm0.57)}$ & 0.79 & $<10^{-4}$ & 0.39\\
   & BL Lacs & $ L_{\rm syn,52}=10^{(-9.51\pm0.63)}\Gamma^{(2.27\pm0.51)}$ &0.69 & $2.14\times10^{-4}$ & 0.64\\
   & NLS1 & $ L_{\rm syn,52}=10^{(-9.71\pm0.37)}\Gamma^{(3.89\pm0.41)}$ & 0.93 & $<10^{-4}$ & 0.34\\
     & AGN+GRB & $L_{\rm syn,52}=10^{(-10.88\pm0.32)}\Gamma^{(4.64\pm0.20)}$ & 0.92 & $<10^{-4}$ & 1.20\\
      \hline
        $L_{\rm bol}(\Gamma)$ & FSRQs & $L_{\rm bol,52}=10^{(-9.13\pm0.45)}\Gamma^{(4.58\pm0.38)}$ & 0.93 & $<10^{-4}$ & 0.26\\
    & BL Lacs & $ L_{\rm bol,52}=10^{(-8.57\pm0.70)}\Gamma^{(2.32\pm0.57)}$ & 0.66 & $4.87\times10^{-4}$ & 0.71\\
    & NLS1 & $ L_{\rm bol,52}=10^{(-8.87\pm0.40)}\Gamma^{(4.13\pm0.43)}$ & 0.93 & $<10^{-4}$ & 0.36\\
     & AGN+GRB & $L_{\rm bol,52}=10^{(-8.29\pm0.27)}\Gamma^{(3.20\pm0.17)}$ & 0.89 & $<10^{-4}$ & 1.03\\
   \hline
     $E_{\rm syn,z}(\Gamma)$ & FSRQs & $ E_{\rm syn,z}=10^{(-4.01\pm0.49)}\Gamma^{(-0.03\pm0.41)}$ & 0.02 & 0.94 & -\\
     & BL Lacs & $ E_{\rm syn,z}=10^{(1.34\pm1.17)}\Gamma^{(-1.86\pm0.96)}$ & -0.38 & 0.06 & -\\
     & NLS1& $ E_{\rm syn,z}=10^{(-4.16\pm0.33)}\Gamma^{(-0.34\pm0.36)}$ &   -0.24 & 0.34 & -\\
   \hline
     $L_{\rm syn}(E_{\rm syn,z})$ & FSRQs & $ L_{\rm syn,52}=10^{(-1.83\pm1.82)}E_{\rm syn,z}^{(0.82\pm0.46)}$  &0.36 & 0.09& -\\
    & BL Lacs & $ L_{\rm syn,52}=10^{(-6.84\pm0.22)}E_{\rm syn,z}^{(-0.05\pm0.14)}$ & -0.08 & 0.71 & -\\
     & NLS1 & $L_{\rm syn,52}=10^{(-7.06\pm3.39)}E_{\rm syn,z}^{(-0.19\pm0.76)}$ & -0.06 &0.81 & -\\
\hline
     $L_{\rm bol}(E_{\rm syn,z})$ & FSRQs & $ L_{\rm bol,52}=10^{-3.03\pm2.25}E_{\rm syn,z}^{0.17\pm0.57}$ & 0.07 &0.76 & -\\
     & BL Lacs & $ L_{\rm bol,52}=10^{-5.92\pm0.23}E_{\rm syn,z}^{0.16\pm0.15}$ & -0.22 & 0.30& -\\
     & NLS1 & $L_{\rm bol,52}=10^{-7.97\pm3.54}E_{\rm syn,z}^{-0.63\pm0.79}$ & -0.20 & 0.44 &-\\
\hline
	\end{tabular}
\\
\end{center}
\end{table}

\begin{table}
%\footnote
\begin{center}
	\caption{Results of our linear regression analysis with a model of $\log L=a+b\log E+c\log \Gamma$ in the source frame for the AGNs (or GRBs) in our samples.}
	\label{Tab:triple}
	\begin{tabular}{lllllll}
    \hline\hline
     relations  &  Source &  $\rm Expressions$ & $p^{a}_{\rm F} $ & $r^{b}$  &  $p^{b}$      &      $\Delta^{b}$ \\
 \hline
      $L^{r}_{\rm syn}(E_{\rm syn,z},\Gamma)$  & FSRQs &   $L^{r}_{\rm syn, 52}=10^{(-5.86\pm1.78)}E_{\rm syn,z}^{(0.79\pm0.26)}\Gamma^{(3.31\pm0.48)}$ &$1.36\times10^{-6}$  &0.86& $<10^{-4}$ & $0.27$\\
     & BL Lacs &   $L^{r}_{\rm syn, 52}=10^{(-9.71\pm0.64)}E_{\rm syn,z}^{(0.15\pm0.11)}\Gamma^{(2.54\pm0.55)}$ & $ 5.62\times10^{-4}$  &0.71& $<10^{-4}$ & $0.44$\\
     & NLS1 &   $L^{r}_{\rm syn, 52}=10^{(-7.58\pm1.19)}E_{\rm syn,z}^{(0.51\pm0.27)}\Gamma^{(4.07\pm0.39)}$  &  $ 2.39\times10^{-7}$  &0.94& $<10^{-4}$ & $0.28$\\
     & FSRQs+NLS1 &   $L^{r}_{\rm syn, 52}=10^{(-7.40\pm0.77)}E_{\rm syn,z}^{(0.45\pm0.15)}\Gamma^{(3.50\pm0.25)}$ &  0 &0.94& $<10^{-4}$ & $0.30$\\
     \hline
        $L^{r}_{\rm bol}(E_{\rm syn,z},\Gamma)$  &  FSRQs &   $L^{r}_{\rm bol, 52}=10^{-8.59\pm0.94}E_{\rm syn,z}^{0.13\pm0.20}\Gamma^{4.57\pm0.38}$& $8.11\times 10^{-10}$ &0.94& $<10^{-4}$ & $0.23$\\
    &   BL Lacs &   $L^{r}_{\rm bol, 52}=10^{-8.60\pm0.73}E_{\rm syn,z}^{0.03\pm0.13}\Gamma^{2.37\pm0.63}$ &$2.6\times 10^{-3}$&0.67& $4.77\times 10^{-4}$ & $0.47$\\
     &  NLS1 &   $L^{r}_{\rm bol, 52}=10^{-8.51\pm1.40}E_{\rm syn,z}^{0.09\pm0.32}\Gamma^{4.17\pm0.46}$ &$1.03\times 10^{-6}$&0.93 &$<10^{-4}$ & $0.33$\\
     & FSRQs+NLS1 &   $L^{r}_{\rm bol, 52}=10^{(-8.16\pm0.71)}E_{\rm syn,z}^{(0.22\pm0.14)}\Gamma^{(4.48\pm0.23)}$ &  0 &0.96& $<10^{-4}$ & $0.29$\\
 \hline
	\end{tabular}
\\
\end{center}
$^{a}$ $p_{\rm F}$ is the probability of the F-test for our linear regression analysis results.\\
$^{b}$ $r$ and $p$ are the linear correlation coefficient and chance probability derived from the Spearman correlation analysis for each pairs of $L_{\rm r}$ and $L$. $\Delta$ is the $1\sigma$ dispersion of the pair correlation.
\end{table}

%\clearpage

%\clearpage
%\clearpage

\section{Conclusion and Discussion}
We have presented analysis on the Doppler boosting effect on the observed luminosity and photon energy in AGNs and GRBs. Our analysis show $L_{\rm syn} \propto \Gamma^{2.27\sim 4}$ for the individual samples of the FSRQs, BL Lacs, and NLS1 galaxies. Similar relations are also found for $L_{\rm bol}(\Gamma)$. They are also globally follow the same $L-\Gamma$ relation as $L\propto \Gamma^{4.64\pm0.20}$ together with the GRBs. A tight relation $L_{\rm syn}\propto E^{0.45\pm0.15}_{\rm syn,z}\Gamma^{3.50\pm0.25}$ is found in the combined sample of FSRQs and NLS1 galaxies. This relation is different from that derived from the GRB sample.

\begin{figure}[htbp]
\centering
\includegraphics[width=0.45\textwidth, angle=0]{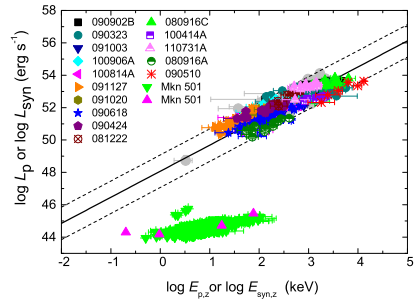}
\caption{$L_{\rm syn}-E_{\rm syn,z}$ relation of Mkn 501 in different outbursts in comparison to the $L_p-E_{\rm p, z}$ relation within individual GRBs in different time slices. The data of Mkn 501 are taken from \citealp{Zhang+etal+2013b} (the pink triangles) and \citealp{Wang+etal+2019} (the green triangles). The GRB data are taken from \citealp{Lu+etal+2012}.}
\label{Mkn 501}
\end{figure}

As shown in \cite{Lyu+etal+2014}, the different distributions of GRBs and blazars in the $L_{\rm syn}({\rm or\ }L_{\rm iso})-E_{\rm syn}({\rm or\ } E_{\rm syn})$ plane may be due to both different radiation physics and jet environments. In addition, blazars have violent variability and a tentative flux-$E_{\rm syn}$ positive correlation is found in some blazars, such as 3C 279 and Mkn 501 (e.g. \citealp{Zhang+etal+2013b}; \citealp{Wang+etal+2019}). Taking the $L_{\rm syn}$, $E_{\rm syn}$, and $\delta$ values of Mkn 501 from \cite{Zhang+etal+2013b}, we show Mkn 501 in four bright outbursts in Figure ~\ref{L-E-Gamma-GRB}. One can find that it deviates the $L_{\rm p,z}-E_{\rm p,z}-\Gamma_{0}$ relation of GRBs in these outbursts. \cite{Wang+etal+2019} studied the $L_{\rm p}$-$E_{\rm p}$ relation for the Mkn 501 in different outbursts in a broad temporal coverage. They found that a weak $L_{\rm syn}-E_{\rm syn, z}$ correlations in some outbursts. We further examine whether it follows the the $L_{\rm iso}-E_p$ relation within individual GRBs (\citealp{Liang+etal+2004}; \citealp{Lu+etal+2012}) in these outbursts. As shown in Figure ~\ref{Mkn 501}, it still does not follow the $L_{\rm iso}-E_p$ relation of GRBs. 

The observed luminosity is boosted by a factor of $\Gamma^{p}$, where $p=2+\beta$ for a continuous jet, $p=3+\beta$ for a moving sphere, and $\beta$ is spectral index of the synchrotron radiation emission (\citealt{Ghisellini+etal+1993}). Our analysis results for the different sub-classes of AGNs are consistent with the prediction of the Doppler boosting effect. The dependence of $L$ on $\Gamma$ in the $L-E_{\rm syn, p}-\Gamma$ relation of the FSRQs and NLS1 galaxies is still consistent with this prediction. However, the dependence of $L$ on $\Gamma$ for the GRBs significantly deviate this prediction. This may be due to the initial Lorentz factor $\Gamma_0$ is not a true representative of the bulk motion of the radiating region. The $\Gamma_0$ values in this analysis are the Lorentz factor of the forward shocked medium derived from the fireball deceleration time (the afterglow onset peak time) based on the standard afterglow model (e.g. \citealt{Sari+etal+1998}). In addition, GRB jets are episodic. They are composed of erratic shells with different initial Lorentz factor since their energy input and baryon matter loaded may be different. Therefore, the $\Gamma_0$ values derived form the afterglow data may not be the true Lorentz factor of the radiating region of the prompt gamma-rays. For example, \cite{Wang+etal+2000} inferred that the initial Lorentz factor of the fireball of GRB 990123 is 1200 and the Lorentz factor at its prompt optical emission peak time is 300.

\cite{Nemmen+etal+2012} illustrated that GRB jet luminosity  is correlated with the jet power, and this relation is consistent with the correlation between jet power and the synchrotron peak luminosity of some AGNs (see also \citealt{Zhang+etal+2013a, Wang+etal+2014}). Note that a substantial fraction of the kinetic energy of the baryons should transferred to a non-thermal population of relativistic electrons through Fermi acceleration in the shock (e.g. \citealt{Meszaros+Rees+1993}). $L_{\rm p}$ is almost proportional to $\Gamma_0$ is within the error of the power-law index in the $L_{\rm p}-E_{\rm p, z}-\Gamma_0$ relation. We suspect that $\Gamma_0$ may be a representative of the kinetic power, or at least the power carried by the radiating electrons, in the radiating region (jet or jet patch). As discussed in \citealt{Lyu+etal+2014}, the different $L-E_{\rm p}$ relation in GRBs and blazars may be resulted from different scenarios of synchrotron radiations. The tight $L_p-E_{\rm p, z}-\Gamma_0$ relation may suggest that the observed gamma-ray luminosity of GRBs depends on the radiation physics and the jet power together.

\begin{acknowledgements}

We appreciate helpful comments from the anonymous referee. We thank J. Zhang, Y. J. Wang, and Y. Q. Xue for providing the data of Mkn 501 and suggestive discussion. This work is supported by the National Natural Science Foundation of China (Grant No.11533003, 11851304, and U1731239),
Guangxi Science Foundation and special funding for Guangxi distinguished professors (2017AD22006).
\end{acknowledgements}

\end{document}